# Analogy of space-time as an elastic medium – Estimation of a creep coefficient of space from space data via the MOND theory and the gravitational lensing effect - the ball cluster and via time data from the GPS effect– comparison, discussion and implication of the results for dark matter and Einstein's field equation

David Izabel[1]

1 Engineer INSA Rennes and mechanical professor, Paris, France

University of Strasbourg CNRS France

d.izabel@aliceadsl.fr

**Summary**

After recalling the principles that allow space-time to be considered by analogy as an elastic medium, we show how the modified gravity according to the MOND theory concerning the anomaly of the velocities of stars at the periphery of galaxies can be seen as a creep of space acting on the radius of galaxies that gives a creep coefficient of $\varphi_{space} = \frac{a_0}{a}\frac{\rho_{local}}{\rho_{mean}} - 1$. The values vary between 0.2 and 9 depending on the type of galaxy and density distribution. Considering the gravitational lensing effect of the ball cluster we obtain a creep coefficient $\varphi_{space} = \frac{1-p_v}{p_v}$ with $p_v$ the percentage of visible matter and $p_{DM}$ the percentage of Dark matter from the global mass ($p_v + p_{DM} = 1$. The values vary between 0.66 and 4 for this cluster. This paper therefore raises the question, via these creep coefficients, of the possible granular nature of the vacuum and therefore of space fabric on the one hand and proposes another dark matter-free approach based on the creep of the texture of space to explain gravitational anomalies on the other hand.

**Introduction**

Cosmology is the study of the behavior of space-time constituting the universe as a whole characterized by Einstein's gravitational field equation **[1], [2]** on the one hand using the Friedmann-Lemaitre-Robertson-Walker metric on the other hand. However, some gravitational phenomena can only be explained by using a mysterious mass or dark matter.

The first phenomenon is the speed of stars on the outskirts of galaxies. F. Zwicky in the 1930s suggested that a certain dark matter could explain the discrepancy between the velocities of stars predicted according to Newton's theory, which are supposed to decrease as we move away from the center of galaxies, and those actually measured where the rather high speed remains almost constant (see Figure 1).

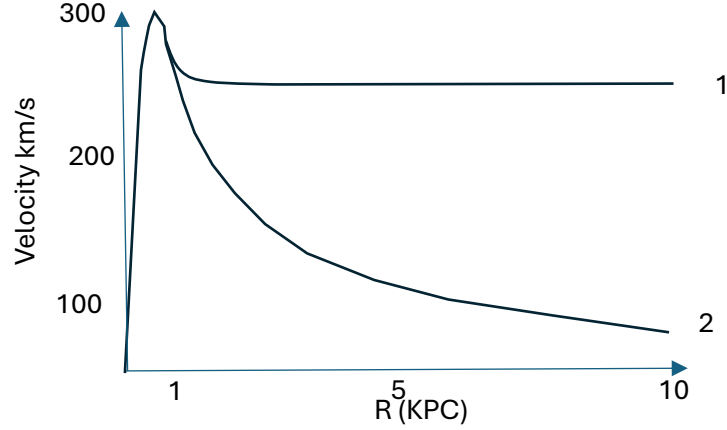

**Figure 1: Velocity of stars in galaxies of radius R according to Newton's theory (2) and observed (1)**

But for more than fifty years, the mysterious dark matter particles have been untraceable, regardless of the detectors and the sensitivity of the detectors. The ontological approach therefore fails for the moment. This leaves the legislative approach, which consists of changing the law of gravitation, at least in part, in certain acceleration ranges. This is the goal of the modified gravity theory developed by Mordehai Milgrom **[3]** and supplemented by other scientists **[4]**.

The second phenomenon occurs at the level of the cosmic microwave background where dark matter is again needed to create the small variations in density of the original plasma that led to galaxies today.

The third is at the level of gravitational lensing, where the mass of galaxy clusters alone is insufficient to explain the amplitude of the observed light distortions.

In this paper, we propose to approach the problem of dark matter through the prism of the space-time analogy with an elastic medium, in the specific case of the anomaly of the velocities of stars at the periphery of galaxies for space, and to see the consequences and interpretations for time based on the feedback in this field, namely the GPS shift and the desynchronization of clocks in the Hafele and Keating experiment. Considering that Einstein's equation replaces gravitation as a force with a deformation of space-time, it implies modeling spacetime as a deformable elastic medium. This has been done by many authors in subsequent publications **[5]** to **[16].** Like any elastic medium, it is then necessary to define the characteristics of the medium, its Young's modulus between $10^{31}$ and $10^{113}$ Pa according to publications **[5]** to **[7], [10],** its Poisson ratio **[5]** to **[7], [10],** of the order of 1, its coefficient of thermal expansion **[8]** of the order of that of steel and therefore, to be complete, its creep coefficient if the medium is sensitive to this phenomenon, which is well known in mechanics.

The state of the art on the creep of space-time is very limited but not zero. We found only two papers **[17]** and **[21]**. In the first **[17]**, the author shows that the equation that manages the propagation of dislocations can be seen as a creep of atoms in a crystal and that a similar expression can be written in cosmology for the vacuum from the cosmological constant $\Lambda$. Thus, in formula 24 of **[17],** the dislocation density ρ is a function of the volume density of the elastic energy $U_{\sigma i}$ of the solid crystal:

$$\rho = \frac{4(1+\nu)}{\alpha^2 M^2 \mu_m b^2} U_{\sigma i} \quad (1)$$

α is a constant that characterizes the interaction of dislocations and depends on the geometric arrangement of the dislocation structure ($\alpha \cong 0.5$):

$$\mu_m = \frac{E}{2(1+\nu)} \quad (2)$$

$\mu_m$ is the usual shear modulus

M is a geometric factor called the Taylor factor,

b is the modulus of the Burgers vector,

$\nu$ is the Poisson coefficient.

In formula 31 of **[17]**, the cosmological constant $\Lambda$ is a function of the density of the energy of the vacuum $U_v$ on the other hand:

$$\Lambda = \frac{8\pi G}{c^4} \times U_v \quad (3)$$

Thus, $U_\sigma$ for dislocation creep plays the role of $U_v$ in relativistic cosmology. This point, based on **[17]** is essential for making the case that space can be assumed as a crystalline structure and therefore is subject to creep.

In **[17]**, the author seriously considers the hypothesis that the vacuum is a crystalline structure subject to everything that occurs in this type of media, namely dislocations and creep. Therefore, considering a possible creep of the vacuum associated with the nature of space and time within space-time becomes conceivable if it has a granular and/or crystalline structure. Finally, in special relativity, time is associated with c which squared gives it the dimension of a squared length with an opposite sign to the Pythagorean sign of space. By definition, any creep characterizes, depending on the nature of a material, the ability of the latter to deform under constant load over time and therefore to vary in length by displacements over time under a constant load without adding or modifying the initial mass that deforms and curves it. Therefore, considering a creep of space from space and time data becomes possible in general relativity.

At the same time, feedback from general relativity and cosmology shows that space-time is clearly subject to more gravitational effects than the visible mass alone allows us to envisage in view of what we have stated in the previous paragraphs. However, according to Einstein, gravitation is in fact a geometric phenomenon.

So, the first approach to explain this observation is to solve the paradox (more gravitation than mass to cause it) by fictitiously increasing the mass that generates this gravitation. Since we don't see it, it has been called dark matter. The problem is that it is not made of traditional matter, is invisible, interacts with classical matter in the sense of quantum field theory and standard model, has a major gravitational effect and remains to this day absolutely undetectable other than by the distortions of space-time that it creates.

The second approach, as we have seen, is to modify Newton's law, as Mordechai Milgrom did. But in this case, why would nature need two laws of dynamics? So, doesn't Ockham's razor apply to space-time?

It is from this point of view that the second paper **[21]** proposes to reinterpret this modification of Newton's law by considering that the anomalous gravitational effects currently attributed to dark matter can alternatively be explained as a manifestation of the inherent structure of space at the galactic level by an effect of length scales. Specifically, the authors show that the inherent curvature of space amplifies the gravity of ordinary matter such that the effect resembles the presence of the hypothetical hidden mass. Their study is conducted in a context of low gravity, quasi-static conditions, and spherically symmetrical configuration, and exploits the Cosmic Fabric model of space **[5]**. It is in this same philosophy that this paper is placed, considering that it is the very texture of the fabric of space that creep under the permanent effect of the internal constraints generated by ordinary matter.

Other approaches are based on in binding of cosmological structures by massless topological defects **[44]**. So, no need of dark matter at all.

Finally, our innovative approach, in link with **[21]**, is to rely on the analogy of the elastic medium and the well-known behavior of materials such as concrete or polyurethane, for example. The creep we described above. Indeed, the data on the problem is clear. We observe a gravitational effect greater than there is mass to cause it or translate into an Einsteinian approach an amplification of the distortions of space-time that do not correlate with the visible mass that causes it, which remains constant (visible matter). The only phenomenon in mechanics that allows deformations to be amplified over time without modifying the load is the creep of the medium or material that constitutes the structure itself. In a way, when the deformation under self-weight of a concrete slab due to the loss of water during drying or of a polyurethane-core sandwich panel due to the reorganization of the polyurethane polygonal cells increases over time, it is as if a dark mass (rather transparent, invisible) which would charge it giving more deformation over time than the initial one under its mass alone. Seen from the outside, one would therefore have the impression that this invisible mass loads the structure more than there is visible mass, since the deformations increase slowly over time compared to the initial deformation under the sole mass of the structure.

We now need to understand how physically a space creep could be in action to replicate the effects of dark matter, which accounts for about 30% of the elements at play in the universe, the rest being dark energy around 65% or visible matter 5%. Let us first recall how creep works and manifests itself on Earth with well-known materials.

Creep is a physical phenomenon that occurs in materials when they are subjected to constant stress over a long period of time. This process leads to progressive deformation, even if the stress remains unchanged. On Earth, this phenomenon is particularly observed in certain materials under specific conditions, including metals, plastics, rocks, and some ceramics, at high temperatures or under high forces or for reconstituted materials such as concrete or polyurethane.

In the case of metals, creep occurs mainly at temperatures close to 30% to 50% of their melting temperature. For example, in aircraft turbines or high-performance engines, metals subjected to high temperatures can gradually deform under mechanical stress. In their cases, atomic scattering is the main cause of high-temperature creeping. Under the effect of heat, the atoms in the metal structure have more energy and can move, thus causing the material to deform.

In the case of plastics and polymers, creeps are common at ambient or slightly elevated temperatures. For example, plastic materials that are subject to static stress, such as pressure pipes or plastic objects that support constant weights, can deform over time. In their cases, creep is due to the reorganization of polymer chains in the material, where the molecules gradually readjust under the load.

In the case of geological materials, such as rocks and ice, creep is observed over long time scales. For example, the creep of ice in glaciers or that of rocks in the Earth's crust under strong tectonic pressures. In their cases, creep is due to the internal reorganization of mineral or molecular grains under the effect of internal forces in the Earth. In geology, this can take millennia, as in the case of tectonic movements.

In the case of composite materials, water loss in the case of concrete is the source of creeping. see "Creep and relaxation Poisson's ratio: Back to the foundations of linear viscoelasticity. Application to concrete » **[40]**

Finally, in ceramics, which are generally very rigid, they can be creeped at very high temperatures, for example in reactors or combustion systems where the temperature is extreme. In their case, the creep mechanism is often due to diffusion processes similar to those seen in metals.

In summary, the parameters that influence creep are therefore high temperatures, prolonged stresses and the nature of the media, grains, polymer with deformable geometric shape connected with the Poisson's Ratio **[40]**.

We must therefore see to what extent the space-time medium could be affected by this phenomenon.

First of all, all research considers an extremely fine texture of Planck's size. Quantum gravity considers granularity of spacetime **[39] [17]**, **[32]**, In **[5]** hyperspace sheets have the characteristics of a fabric or connected fibers with a Poisson ratio of 1. In **[5]** the authors cite two publications **[30]** and **[31]** where Poisson's coefficients of 1 and where the texture of the medium is fibrous or in the form of a polygon that deforms. So, Poisson's ratio of 1 implies somewhere a fine-grained or fiber-shaped or polygonal deformable structure, all of which are polymer-like creep source parameters.

If we now look at the Young's modulus of the space fabric, it turns out to be very high and greater than $10^{20}$ times that of steel, in which case only temperature could make it flow. Here too between the hot zones of the cosmic web, temperatures of several million degrees could be enough to generate metal-like creep.

In conclusion, Poisson's ratio of 1 is a strong indication of the potential nature of the elastic medium that in the analogy could flow, i.e. fine grains or fibers or polygonal meshes of a fabric that can all be sensitive to creeping.

Creep is therefore the result of mechanisms at the atomic or molecular scale that allow the particles of the material to gradually rearrange themselves under the effect of constant stress and a given temperature. The space-time medium being charged by the entire cosmic web for more than 13.7 billion years, we can therefore understand if the nature of the elastic medium allows it to undergo a creep effect, i.e. constant deformations, therefore constant additional equivalent gravitation, without adding mass. Consequently, Granular materials are subject to creep in compression **[43]** or shear **[24]**.

We therefore propose considering an identical phenomenon for rotating space-time in the case of a galaxy (the quantity of stars inside in rotation load the space fabric in continue for long time) or deformed by a cluster of galaxies during gravitational lensing or even during the hot big bang. We will therefore see in this article whether finally the modification of Newtonian dynamics proposed by the MOND's theory, which works but does not go in the direction of simplification, does not amount to considering a creep coefficient to apply $\kappa$ Einstein's constant which characterizes the flexibility of space-time itself and can therefore be correlated from the point of view of the analogy of space-time to a correction as a function of the time of the Young's modulus of the latter. In fact, we can consider that the creep has these two effects: (1) Affect the constitutive properties of space resulting in an effective gravitational constant that is different from the conventional one, and (2) it modifies the geometry of space resulting in additional curvature. Then point out that we are focusing on (1) and that the consequences of (2) are covered in Ref **[21]**. Finally, if we modify the constant $\kappa$ to incorporate these creep effects, then it implies that not only space but also time is impacted by this effect. We will therefore try to establish a creep coefficient also for time and compare it with the one obtained for space with the MOND theory and gravitational lensing (ball 's cluster) to see if all these values are consistent.

## 2. Method

The following methodology has been followed in this article.

1) Reminder of the equations of general relativity and the analogy with the mechanics of continuous media,
2) Reminder of the notions of creep in mechanics,
3) Reminder of the theory and the MOND equations associated with the speed of stars on the periphery of galaxies,
4) Reexpression of the MOND equations and Newtonian gravitation in the case of the problem of the velocities of stars at the periphery of galaxies in terms of variation of equivalent geometric radii of the latter,
5) Integration of creep in general and Newtonian relativity on the gravitational constant G,
6) Evaluation of the creep coefficient of space $\varphi_{space}$,
7) Evaluation of the creep coefficient of space from another characteristic source, the ball cluster,
8) Expression of a creep coefficient in relation to time fluctuations and data in a weak gravitational field,
9) Evaluation of the creep coefficient of time from GPS within the vicinity of the Earth,
10) Discussion, comparison of the results obtained with other properties of space-time compared with other phenomena and other theories.

## 3. Reminder of the equations of general relativity and the connections with the analogy of the elastic medium

The law of gravitation according to general relativity is written with $R_{\mu\nu}$ the Ricci tensor, $g_{\mu\nu}$ the metric tensor, R the scalar curvature, G the gravitational constant, and c the speed of light:

$$R_{\mu\nu} - \frac{1}{2} g_{\mu\nu} R = \frac{8\pi G}{c^4} T_{\mu\nu} \quad (4)$$

In a weak field it becomes:

$$\Box \bar{h}_{\mu\nu} = \frac{16\pi G}{c^4} T_{\mu\nu} \quad (5)$$

With:

$$\bar{h}_{\mu\nu} = h_{\mu\nu} + \frac{1}{2} \eta_{\mu\nu} \bar{h} \quad (6)$$

And $\bar{h}$ the trace of $\bar{h}_{\mu\nu}$

We have shown in **[5]** to **[7]** that the perturbation tensor of the metric $h_{\mu\nu}$ in weak gravitational field is equivalent to a strain tensor in 4 dimensions $\varepsilon_{\mu\nu}$ by a factor of 2:

$$h_{\mu\nu} = 2\varepsilon_{\mu\nu} \quad (7)$$

And it has been shown **[6]** that the stress-energy tensor $T_{\mu\nu}$ is equivalent to a stress tensor $\sigma_{ij}$ with the velocity vectors $v_i$ and $v_j$, the four velocity vectors $u_\mu$ and $u_\nu$:

$$\sigma_{ij} = \rho v_i v_j \quad (8)$$

$$T_{\mu\nu} = \rho u_\mu u_\nu \quad (9)$$

Finally, it is well known **[5] to [7]** that Einstein's constant κ is equivalent to the flexibility of spacetime for a unit surface in N⁻¹ with Y the Young's modulus of spacetime and S a unit surface:

$$\kappa = \frac{8\pi G}{c^4} \rightarrow \frac{1}{YS} \quad (10)$$

This is why Einstein's equation is often considered to be Hooke's law with ε a deformation, E the Young's modulus of the elastic material and σ a stress.

$$\varepsilon = \frac{1}{E}\sigma \quad (11)$$

This leads according to **[5]** to **[8] [10], [18]** to consider that the gravitational constant G is related to the Young's modulus of space-time via a frequency f of eigen vibration to the vacuum and ρ a density of the vacuum:

$$G = \frac{\pi f^2}{\rho} \quad (12)$$

Which becomes with $Y = \rho c^2$ the expression that presents the propagation of compression waves in an elastic medium:

$$G = \frac{\pi f^2 c^2}{Y} \quad (13)$$

Let be a κ mechanized constant **[6], [7], [17]**:

$$\kappa = \frac{8\pi G}{c^4} \rightarrow \frac{1}{YS} \quad (14)$$

Or:

$$\kappa = \frac{8\pi^2 f^2 c^2}{Yc^4} = \frac{8\pi^2 f^2}{Yc^2} \quad (15)$$

So, $\kappa$ corresponds to the flexibility of the unit area (S=1m²) of space-time and is a function of 1/E=1/Y.

**4. Reminder of the concepts of creep in mechanics**

If space is modeled as an elastic medium, then this typology of medium could be subject to creep via a creep coefficient φ, if by analogy it is comparable to a granular or crystalline structure. We recall that in quantum gravity we assume that space-time is grainy. In sandwich theory, creep corresponds to the polygonal cells of polyurethane that deform under loading, influencing the shear modulus μ of this core. In concrete steel flooring, for example, the concrete creep is due to water loss and therefore when the bending deformations of the slabs are calculated, the cracked and uncracked concrete sections are calculated with a Young's modulus E that fluctuates as a function of time.

Concretely, in mechanics, creep is introduced by a coefficient φ by varying the Young's modulus or shear modulus in the following way with $E_{st}$ the short-term Young's modulus and $E_{Lt}$ the long-term Young's modulus:

$$E_{Lt} = \frac{E_{St}}{(1+\varphi)} \quad (16)$$

In the case of the shear modulus, we have the short-term shear modulus $\mu_{st}$ and the long-term shear modulus $\mu_{Lt}$ in the same way:

$$\mu_{Lt} = \frac{\mu_{St}}{(1+\varphi)} \quad (17)$$

In both cases, the objective is to increase the deflection of the beam by decreasing the Young's modulus.

For example, in the case of a beam of inertia I on two supports, with a span L, uniformly loaded by p, with a bending creep-sensitive, the vertical deflection is written with φ>0:

Without creep:

$$f_{without\ creep} = \frac{5pL^4}{384EI} \quad (18a)$$

With creep:

$$f_{with\ creep} = \frac{5pL^4}{384EI}(1+\varphi) \quad (18\ b)$$

So, a ratio of deflection:

$$\frac{f_{without\ creep}}{f_{with\ creep}} = \frac{1}{(1+\varphi)} \quad (18\ c)$$

If space-time is subject to creep, the constant κ that represents the flexibility of creep could vary as a function of G that can be traced back to the Young's modulus of spacetime that would vary over time as seen in paragraph 3. We would then obtain for the creep flexibility of space-time:

$$\kappa = \frac{8\pi G}{c^4} \rightarrow \frac{8\pi^2 f^2 c^2}{\frac{Y}{(1+\varphi)}c^4} = \frac{8\pi^2 f^2(1+\varphi)}{Yc^2} \quad (19a)$$

Or with a gravity modified by creep effect:

$$G_{Lt} = \frac{\pi f^2 c^2}{Y}(1+\varphi) = G(1+\varphi) \quad (19b)$$

A new κ expression corrected by creep is therefore written:

$$\kappa = \frac{8\pi^2 f^2 c^2}{\frac{Y}{(1+\varphi)}c^4} = \frac{8\pi G(1+\varphi)}{c^4} \quad (20)$$

Thus, with φ > 0, the flexibility of space will increase, and the rigidity will decrease. So, the radius of curvature and deflection will increase and the ratio 1/R² will decrease.

## 5. Reminder of MOND's theory and equations

It is well known that the curve of the rotation speeds of stars in spiral galaxies does not follow the classical Newton's law. The velocities remain radially constant as one moves away from the center of the galaxy (see Figure 1). The MOND Theory (Modified Newton Dynamic) proposed by Mordehai Milgrom [3] [4] makes it possible to reproduce this velocity curve precisely without using dark matter but does not explain why this is so.

Thus, the expression according to Newton's law with M the mass of the galaxy and r the radius of calculation of the speed $v$ of the stars with respect to its center and G the gravitational constant is written:

$$v = \sqrt{\frac{GM}{r}} \quad (21)$$

And dimensional analysis indeed produces units of a velocity:

$$\left(\frac{\frac{m^3}{kgs^2}kg}{m}\right)^{1/2} = \frac{m}{s} \quad (22)$$

This expression of velocities $v$ according to Newton is replaced according to **[3] [4]** by a constant value that no longer depends on the radius r of the galaxy (see demonstration below):

$$v = \sqrt[4]{GMa_0} \quad (23)$$

The equation of dimensions allows us to verify that it is indeed a velocity:

$$\left(\frac{m^3}{kgs^2}kg\frac{m}{s^2}\right)^{1/4} = \frac{m}{s} \quad (24)$$

With $a_0$ a new constant in physics **[19]**:

$$a_0 = 1.2 \times 10^{-10} m/s^2 \quad (25)$$

We can therefore see that according to this MOND approach, the gravitational constant is corrected by a factor $a_0$ that we will somehow transform to an equivalent concept within the domain of Mechanics to deduce a creep coefficient $\varphi_{space}$ for the geometry of space.

MOND's new law of dynamics **[3], [4] and [20],** i.e. the force F of gravitation in the case of galaxies, is written with $\mu$ a function of $\frac{a}{a_0}$ which modifies the classical acceleration a:

$$F = \frac{GMm}{r^2} = m\mu\left(\frac{a}{a_0}\right)a \quad (26)$$

Or according to **[4] [20]**:

$$a = \sqrt{F\frac{a_0}{m}} \quad (27)$$

This makes it possible to write by simplifying by m:

$$F = \frac{GM}{r^2} = \mu\left(\frac{a}{a_0}\right)a \quad (28)$$

When the radius r of the galaxy is very large, a is smaller than $a_0$ and we therefore obtain $\mu\left(\frac{a}{a_0}\right) = \frac{a}{a_0}$ which give:

$$\frac{GM}{r^2} = \frac{a^2}{a_0} \quad (29)$$

With the radial acceleration which is classically equal to:

$$a = \frac{v^2}{r} \quad (30)$$

Let us transfer this acceleration to MOND's expression:

$$\frac{GM}{r^2} = \frac{v^4}{r^2 a_0} \quad (31)$$

Let be the constant speed of the stars in the galaxy:

$$v^4 = GMa_0 \quad (32)$$

Let be the final expression of the speed of stars in a galaxy **[19] [20].**

$$v = \sqrt[4]{GMa_0} \quad (33)$$

It can therefore be seen that MOND's solution makes it possible to give a constant velocity of the stars on the periphery of galaxies independent of the radius r of the latter, which corresponds well to what has been measured experimentally.

From a mechanical point of view, it is as if the galaxy were caught in a space disk constituting a rotating elastic medium in which all the stars trapped in it are driven at the same speed.

**6. Reexpression of the MOND equations in terms of geometric variations of the radius r of an elastic disk constituting the galaxy and comparison with the same radius in the case of the classical Newtonian approach**

From the expression of the velocities established in the MOND theory, we obtain the expression of the velocity squared of the stars:

$$v^2 = \sqrt{GMa_0} \quad (34)$$

With $a = \frac{v^2}{r}$ for the acceleration and replacing $v^2$ by this acceleration a in the above expression of the velocities of the stars in the galaxy according to the MOND's theory, we obtain an equivalent radius of the galaxy that we denominate $r_{mond}$:

$$ar = \sqrt{GMa_0} \quad (35)$$

And we deduce an equivalent radius of the galaxy according to the MOND theory:

$$r_{mond} = \sqrt{\frac{GM}{a}\frac{a_0}{a}} \quad (36)$$

We can also reformulate the radius of the galaxy in the case of classical Newtonian gravitation from $F = \frac{GMm}{r^2} = ma$. Let with $a = \frac{GM}{r^2}$, we get:

$$r_{Newton} = \sqrt{\frac{GM}{a}} \quad (37)$$

So, finally the MOND theory can be seen as a modification of the radius of the galaxy's disk by a ratio of acceleration a of it with respect to a reference acceleration $a_0$:

$$r_{Mond} = \sqrt{\frac{GM}{a}\frac{a_0}{a}} = r_{Newton}\sqrt{\frac{a_0}{a}} \quad (38)$$

By comparing the expression of the equivalent radius of the galaxy according to the MOND theory with that of the radius of the galaxy according to Newton member by member, it naturally comes:

$$r_{Mond} = \sqrt{\frac{M}{a}\left(G\frac{a_0}{a}\right)} = r_{Newton}\sqrt{\frac{a_0}{a}} \quad (39)$$

Or:

$$r_{Mond}^2 = \frac{M}{a}\left(G\frac{a_0}{a}\right) = r_{Newton}^2 \frac{a_0}{a} \quad (40)$$

So, if a is smaller than $a_0$ (Initial assumptions taken to simplify $\mu\left(\frac{a}{a_0}\right)$), the equivalent radius of the galaxy according to the MOND theory becomes larger. It is as if the radius of the galaxy in the context of the analogy of the elastic medium had creeped! The analogy of the elastic medium gives a mechanical visualization of the structures of the modification of Newton's dynamics so that the velocities of the stars measured in the galaxy correspond to that of the MOND theory.

This constant $a_0$ represents, according to Milgrom, I quote: "It is roughly the acceleration that will take an object from rest to the speed of light in the lifetime of the universe. It is also of the order of the recently discovered acceleration of the universe **[29]**". We can therefore see an increase in the creep of the universe over time in the sense of the analogy of the elastic medium.

## 7. Integration of creep in general relativity and in Newtonian gravitation in the framework of the analogy of the elastic medium

### 7.1 Expression of the creep coefficient linked to space

The mass M and acceleration a of the galaxy being known, we can therefore postulate that the ratio $\frac{a_0}{a}$ affects the gravitational constant as seen in the $r_{mond}$ formulas above. Moreover, we have seen that the constant κ in the framework of elastic analogy is related to the flexibility of space-time in 1/E = 1/Y and therefore with creep in $1/\left(\frac{E}{1+\varphi}\right)$. Using the analogy of the elastic medium, we can therefore compare the value of the short-term gravitational constant with a long-term gravitational constant corrected by a creep effect associated with its Young's modulus Y via a coefficient $(1 + \varphi)$. We therefore obtain by comparing $G$ long-term denoted $G_{Lt}$ with short-term G:

$$G_{Lt} = \frac{\pi f^2 c^2}{Y}(1 + \varphi) = G(1 + \varphi) \quad (41)$$

The relationship between Newton's short-term constant G and long-term $G_{LT}$:

$$G_{Lt} = G(1 + \varphi) \quad (42)$$

We therefore postulate that the MOND's theory effect on the equivalent radii of the galaxy is due to a variation of G by the creep effect of space (since G is associated with the Young's modulus Y in the elastic medium analogy):

$$G(1 + \varphi) = \left(G\frac{a_0}{a}\right) \quad (43)$$

Let us consider the following relation which allows us to evaluate the creep coefficient of space considered by analogy as an equivalent elastic medium:

$$\frac{a_0}{a} = (1 + \varphi) \quad (44)$$

Let the following expression of the creep coefficient of space which, according to the MOND theory seen under the analogy of the elastic medium, implies a variation in the radius of the galaxy.

$$\boldsymbol{\varphi_{space} = \frac{a_0}{a} - 1} \quad \textbf{(45a)}$$

### 7.2. Numerical evaluation of the creep coefficient of the equivalent elastic medium from the MOND's equations explaining the almost constant rotation of stars at the periphery of galaxies

From the above expression, we can proceed to the numerical application to evaluate the value of the creep coefficient φ. We will use for that the data given in publication **[21]**.

For the numerical application we have with r in year light and M in n solar masses:

$$a_0 = 1.2 \times 10^{-10} m/s^2$$

$$a_{(n,r)} = \frac{\sqrt{GMa_0}}{r} = \frac{\sqrt{6.674 \times 10^{-11} \times n \times 10^9 \times 1.98 \times 10^{30} \times 1.2 \times 10^{-10}}}{r \times 1000 \times 365 \times 24 \times 60 \times 60 \times 300000000}$$

Hence the creep coefficient related to space:

$$\varphi_{space} = \frac{1.2 \times 10^{-10}}{a_{(n,r)}} - 1 \quad (45b)$$

In table 1 below, concerning several galaxies, we are looking at the effects of creep as it relates to how it changes the effective gravitational constant and without considering the additional geometric effects of the creep.

| Case | Galaxy | Type of galaxy | Mass M [$10^9$] solar mass) n | Radius r [$10^3$] light years) | $\varphi_{space} = \frac{1.2 \times 10^{-10}}{a_{(n,r)}} - 1$ |
|---|---|---|---|---|---|
| **1** | **Milky way With all the 160 globular cluster around milky way [42]** | **Barred spiral** S(B)bc I-II | **1000 at 1500 1500** | **50 100 520 [42]** | **-0.54922192 -0.26388247 2.827** |
| **2** | **LMC (*Large Magellanic Cloud*)** | **Dwarf spiral** SB(s) m | **10** | **7** | **-0.36891068** |
| **3** | **SMC** | **Dwarf spiral** SB(s) m | **7** | **3.5** | **-0.622852** |
| **4** | **Andromeda** | **Spiral** SA(s)b Interaction with M32 and M110 | **1000** | **110** | **-0.00828822** |
| 5 | M33 | **Spiral** SA(s) cd | 50 | 30 | 0.209564525 |
| 6 | Pinwheel **M101 (NGC 5457)** | **Spiral intermediate** S<u>A</u> B(rs)cd | 100 | 85 | 1.423325287 |
| **7** | **Wirlpool M51 (NGC 5194** | **Spiral** SA(s)bc pec | **160** | **30** | **-0.32383287** |
| 8 | SunFlower messier 63 | **Spiral** SAbc | 140 | 49 | 0.180660002 |
| **9** | **M77** | **Spiral** (R)SA(rs)b Sb/PSb | **1000** | **85** | **-0.23367726** |
| 10 | Condor NGC 6872 | **Barred spiral is interacting with the lenticular galaxy IC 4970**, SB(s)b pec | 100 | 261 | 6.441034116 |
| 11 | Carwheel | **Lenticular** | 4 | 75 | 9.691140972 |
| 12 | Malin 1 ( **PGC 42102**) | **Galaxy geante** | 1000 | 325 | 1.930057542 |
| 13 | **Phonix Cluster** | **Cluster of galaxy** | **2000000** | **550** | **-0.88912325** |
| 14 | **GN-z11 (the furthest in 2016 at 13.4 Year light)** | **Unknown** | **1** | **1.5** | **-0.57235436** |
| Sb-like galaxies can be distinguished from Sc galaxies by a less open spiral structure and a more prominent bulge. Sa-type galaxies have an even less open spiral structure and an even more prominent bulge. Presence of a bar (SB) and open arms with a weak bulb (SBc). | | | | | |

**Table 1 : Estimation of the creep coefficient from MOND approach for several galaxies based on G affected by creeping**

**Discussion of the results obtained**

We see that certain values are positive (half of the cases), that is logical but that also for the other half the values are negative. For memory, creep has these two effects: (1) it affects the constitutive properties of space, such that the long term value of G changes (analogous to the long term value of the Young's modulus $Y_{Long\ term}$), and (2) it affects the geometry of space, where it may lead to additional curvature.

Since here we are only concerned with (1), it is possible that the negative values indicate that (2) has a significant effect that needs to be taken into account.

For these negative values, we first recall the simplifying assumption that has been adopted, which is that the radius *r* must be very large, for the acceleration *a to* be much smaller than *a₀* and therefore the expression of $\mu(\frac{a}{a_0})$ becomes equal to $\frac{a}{a_0}$. So, for all dwarf galaxies (cases 2 and 3) this is no longer the case because of their smallness. The assumption is no longer true and therefore the above approach does not apply.

Then for all galaxies in clusters or interacting with other galaxies, we are not in the simple case of an isolated MOND galaxy either, so for cases 4 and 13 the method does not apply either.

For the galaxy 13.7 billion light-years away (case 14), the dwarf or non-dwarf nature is unknown.

Cases 9, 7 and 1 therefore remain in negative values.

For case 1, it turns out that the real radius, taking into account globular clusters, is much larger than those commonly established in the first approach, see **[42].** It can be seen that the value of φ becomes positive again when taking into account the peripheral globular clusters.

As cases 7 and 9 are of the same type as case 1, a more accurate knowledge of these globular clusters will probably settle this negative value.

So, there are still values of φ varying from 0.18 to 9.7 such as the one in Table 2 below established from **[21]**. How can such discrepancies be explained? Probably 3other parameters come into play.

- The mass/density distribution that could increase creep in certain areas than in other created globally,
- The time of loading between the center of the galaxy with more gravity than in periphery,
- The temperature that can impact as for the Earth material the creep value (see the publication **[41]** on the possible thermal expansion coefficient of the space time fabric).

Concerning the first parameter, we can see from the reading of Table 1 that several galaxies interact with others, some are barred spiral or not, others are giant or dwarf, therefore of different shapes, which necessarily implies a distribution of non-uniform matter that more or less overloads the space fabric and therefore gives by neglecting this phenomenon large variations in the value of the creep coefficient. In our opinion, in order to refine our model, it would be necessary to calibrate the creep coefficient as a function of the mass density at the place where the creep is calculated $\rho_{local}$ with respect to the mean value $\rho_{mean}$:

$$\varphi_{space} = \frac{a_0}{a} \frac{\rho_{local}}{\rho_{mean}} - 1 \ (45a)$$

Regarding the second parameter, the loading time should influence the creep intensity.

Based on the above database, we see that the Milky Way, which is rather recent, and GN-Z11, which is roughly the age of the universe, lead to the same value of the creep coefficient. Therefore, more data would be needed to confirm or refute this stability over time.

Regarding the third parameter, the effect of temperature, we would need to have the temperature variations of each of the galaxies above to quantify how it would influence the fabric of space-time.

Thus, if we are interested in the shape of galaxies and their mass distribution, we could explain these differences.

To verify the magnitude of our results, we can compare them with those obtained in **[21]** where the authors calculated the variation of the radius of the galaxy with an effect inherent to the fabric of space itself. Thus, the variation in radius of the galaxy expressed in the form of the ratio R/s can be interpreted according to our approach as the direct value of the creep linked to the variation in horizontal arrow of the galaxy in its plane.

We find that the results are consistent with our approach in many cases (see table 2).

| Galaxy | Mass M [$10^9$]solar mass) n | Radius r [$10^3$] light years) | Ratio $\frac{R}{s} = \frac{f_{without\ creep}}{f_{with\ creep}} = \frac{1}{(1+\varphi)}$ see formula (18c) | $\varphi$ |
|---|---|---|---|---|
| Milky way | 1000 | 50 | 0.45 | 1.22222222 |
| LMC | 10 | 7 | 0.63 | 0.58730159 |
| SMC | 7 | 3.5 | 0.38 | 1.63157895 |
| Andromeda | 1000 | 110 | 0.99 | 0.01010101 |
| M33 | 50 | 30 | 1.21 | 0.17355372 |
| Pinwheel | 100 | 85 | 2.42 | 0.58677686 |
| Wirlpool | 160 | 30 | 0.67 | 0.49253731 |
| SunFlower | 140 | 49 | 1.18 | 0.15254237 |
| M77 | 1000 | 85 | 0.76 | 0.31578947 |
| Condor | 100 | 261 | 7.42 | 0.86522911 |
| Carwheel | 4 | 75 | 10.67 | 0.90627929 |
| Malin 1 | 1000 | 325 | 2.92 | 0.65753425 |
| Phonix Cluster | 2000000 | 550 | 0.11 | 8.09090909 |
| GN-z11 | 1 | 1.5 | 0.43 | 1.3255814 |

**Table 2 issued of [21] effect of inherent curvature of space with** $s = \sqrt{\frac{GM}{a_0}}$

In reference **[21]** and table 2 describes a geometric effect of creep (modification of shape and so curvature of the geometric structure of the space fabric) that may be complementary to the effect described in table 1 (intrinsic elastic effect of the nature of the space fabric itself). If $G_{Long\ term} > G$ is considered as causing the Dark matter effect, creep could have caused the extra space "dimples" described in **[21]** that would further amplify gravity.

Additionally, the thermal gradient between the cosmic fabric and galaxies and the vacuum can also create additional curvature that changes the geometry and that can also influence the intrinsic creep of the elastic fabric **[41]**.

At this stage, by this approach we arrive at creep coefficients for space between 0.2 and 9.

## 8) Evaluation of the creep coefficient of space from a representative and documented gravitational lensing effect - the ball cluster

Dark matter also manifests itself through gravitational lensing. We then measure deviations of light that are greater than that which should be measured with the effect of visible matter alone.

As a reminder, the angular deviations by gravitational lensing obtained from general relativity are worth **[34]**:

$$\alpha = \frac{4GM}{c^2 R} \quad (46)$$

R is the minimum distance between the light ray and the massive object, M is the mass of the object, c is the speed of light, G is the gravitational constant.

The galaxy cluster that most highlights the effect of dark matter is the ball cluster.

According to the publication **[35]**, the mass share due to visible mass and dark matter mass varies from 60% near the center of the cluster to 20% at the periphery (Figure 12 **[35]**). So roughly 20% of mass bound to visible matter and 80% mass bound to dark matter.

From this data, we can make an estimate of the associated creep coefficient.

We first write that the percentages of mass bound to visible matter $p_v$ and bound to dark matter $p_{DM}$ are 1:

$$p_v + p_{DM} = 1 \quad (47)$$

Which translates basing on formula 46 into percentage of deflection angles for the visible mass $\alpha_v$ and the dark matter mass $\alpha_{DM}$ given the proportionality between the deflection angle $\alpha$ and the mass M that causes it **[34]** by:

$$\alpha_v + \alpha_{DM} = 1 \quad (48a)$$

The creep can be seen as an increase of the angle due to the visible matter $\alpha_v$ by a factor φ:

$$\alpha_v + \varphi \alpha_v = 1 \quad (48b)$$

$$\alpha_v(1 + \varphi) = 1 \quad (48c)$$

We express how much the total deviation is greater than the deviation related to the visible mass $\alpha_v$ by a factor k. Basing on proportionality between α and M and so between $\alpha_v$ and $p_v$ we obtain:

$$k \alpha_v = k p_v = 1 \quad (49)$$

$$k = \frac{1}{\alpha_v} = \frac{1}{p_v} \quad (50)$$

This amplification is expressed by a creep effect by amplification of the strains of space, basing on (48c) and (49) we obtain:

$$k = 1 + \varphi \quad (51)$$

This therefore implies for the creep coefficient φ:

$$\varphi = k - 1 \quad (52)$$

With the expression of k as a function of the mass percentages and:

$$\varphi_{space} = \frac{1}{\alpha_v} - 1 = \frac{1 - \alpha_v}{\alpha_v} = \frac{1 - p_v}{p_v} \quad (53)$$

**Numerical application**

With the data of **[35]** we obtain:

- For a radius close to the center of the galaxy $p_v = 0.6$ gives $\varphi = 0.66$
- For a radius far from the center of the galaxy gives $p_v = 0.2$ gives $\varphi = 4$

The values are therefore consistent with those obtained from the MOND law close to the center of the galaxy but far from it for large radii.

**9. Expression of a creep coefficient in relation to time fluctuations in weak gravitational fields**

**9.1 Expression of the time creep coefficient**

We have seen in the previous paragraphs that we can replace the effects of dark matter in space in the case of the rotation of galaxies by an equivalent creep of the latter.

But since general relativity implies that in connection with special relativity we must also consider time, we must therefore evaluate from the experiments carried out on the lengthening and shortening of time (effect of gravitation and special relativity around the earth by GPS effect, effect of desynchronization of clocks in reverse motion around the Earth in airplanes **[27] [28]** compared to clocks on Earth) what this creep coefficient would be time.

In the case of space (see T. Damour lecture **[25]** and figure 2) it can be shown that curvature results in a satisfactory variation of angles:

$$\hat{A} + \hat{B} + \hat{C} \approx \pi\left(1 + \frac{2GM}{rc^2}\right) \quad (54)$$

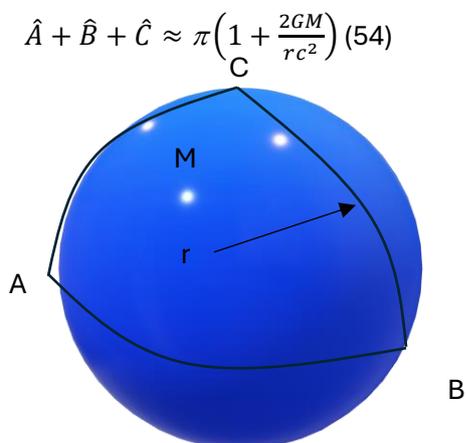

**Figure 2: Effect of the curvature of space on Earth**

General relativity for gravitation and special relativity tells us that time lapse varies. Thus, if the tic – tac translates this duration dt of time, it lengthens and shortens in general relativity according to the following formula for the effect of time between the time on the surface of the Earth and the Earth away from the gravitational field **[25]**:

On the Erath surface we have so:

$$\Delta t_{Surface} = d\tau = \Delta t_\infty \sqrt{1 - \frac{2GM}{rc^2}} \quad (55)$$

The interval is written in the case of the Schwarzschild's metric for the time with $d\tau$ the proper time:

$$d\tau^2 = \left(1 - \frac{2GM}{rc^2}\right) dt^2 \quad (56)$$

That can be written with a limited development in weak field:

$$\frac{dt}{d\tau} = \frac{1}{\sqrt{\left(1-\frac{2GM}{rc^2}\right)}} \approx \frac{1}{1-\frac{GM}{rc^2}} \quad (57)$$

We see that the lapse of time $d\tau < dt$, on the Earth surface, the lapse of time is smaller than at a distance r of the center of gravity mass, so the time (t+dτ) passes less quickly than at a distance r with t+dt, so there is a time dilatation on Earth by gravitation.

$$t + d\tau < t + dt \quad (58)$$

We know from **[5]** that the ration between the proper time and the time of the observer is connected with the volumetric stress $\varepsilon^{3D}$ Matematically, the following variational expression where dτ for the proper time variation and dt for the time variation applies.

$$\frac{d\tau}{dt} = \frac{1}{(1+\varepsilon^{3D})} \quad (59)$$

In publication **[5]** the author explains the link with the time component and the volumetric stress noted $\varepsilon^{3D} = \varepsilon_i^j$. This volumetric stress is also connected with the shear modulus μ by the expression below.

$$\mu = \frac{1}{(1+\varepsilon^{3D})^3} \mu_0 \quad (60)$$

$\mu_0$ is the reference modulus of the undeformed fabric.

We know from **[5]** that the shear modulus μ is connected to the Poisson's ratio that is 1 for space and time in our model based on the Gravitational wave displacement and associated strains measured by the interferometer Ligo/Virgo).

$$\mu = \frac{Y}{2(1+\nu)} \quad (61)$$

We know also with the publications **[30]** and **[31]** (as explanation given in **[5]** about the Poisson ration of 1) that this value of ν implies a particular chape of the space-time as small grains or fibers that can be distorted (fibers, polygonal form grain etc see first part of this paper).

This possible grain form of the time associated with space-time granularity is developed in quantum loop gravity **[39]** and also explained in publication **[17]** and **[32]** "Granular Space–time: The Nature of Time Carlton Frederick of 2022". In this publication, online with a granular time, the author considers that to define time we need not 1 parameter but 2. I quote " a coordinate (t) from minus to plus infinity (or from the big bang to some end of time), and a sequencer (v) (a measure of something coming before or after something else, and the interval between them), an ordering schema as described by H. Reichenbach **[33]** determining the direction and 'speed' of time in link with the arrow of time. Basing on **[33]**, These Schema have the same form that the typical form source of creep in space defined in first part (eg rectangular hexagon following **[32]** a sort of "time polymer shape" we can say) that can be so distorted and thus can create creep of sequence of the time (rather that the time itself). But at the end the variation of time is concerned by this potential sequencer variation that will be study below.

Thus, we arrive at the same hypothesis given as basis the quantum gravity that the granularity of time **[32]** is also the base of the quantum loop gravity **[39]** and interweaving of loops. And this type of fabric is sensible at creep. Space-time is considered as well consider as fabric both for space and time as done in **[5]**. So, we have to find the formula to connect the time creep and the general relativity information of the time elasticity. In weak field we have following **[5]**:

$$1 + 2\varepsilon_{00} \approx g_{00} \quad (62a)$$

And the proper time takes the following expression with convention for the interval (+---):

$$d\tau = g_{00}dt \quad (62b)$$

That becomes:

$$\frac{d\tau}{dt} = (1 + 2\varepsilon_{00}) \quad (63)$$

That we compare with equation (56) reformulated:

$$\frac{d\tau}{dt} = \left(1 - \frac{GM}{rc^2}\right) \quad (64)$$

dτ is smaller than dt.

By comparison with (57) and keeping the first term as $\varepsilon_{00}$ is small:

$$\varepsilon_{00} = -\frac{GM}{2rc^2} \quad (65)$$

So, this strain about time becomes from equation (64):

$$\frac{d\tau - dt}{dt} = -\frac{GM}{rc^2} \quad (66)$$

Taking into account that creep implies an increase of G by a factor (1+φ) we obtain:

$$\frac{d\tau - dt}{dt} = -\frac{GM(1+\varphi)}{rc^2} \quad (67)$$

We can extract φ from (67):

$$-\left(\frac{d\tau - dt}{dt}\right)\frac{rc^2}{GM} - 1 = \varphi \quad (68)$$

Or equivalently:

$$\left(\frac{dt - d\tau}{dt}\right)\frac{rc^2}{GM} - 1 = \frac{\delta t}{\delta t_0}\frac{rc^2}{GM} - 1 = \left(\frac{\frac{\delta t}{\delta t_0}}{\frac{GM}{rc^2}} - 1\right) = \varphi \quad (69)$$

**9.2 Numerical Application to the Effect on Earth in the Case of Time Offset with GPS Satellites**

It is well known that Newtonian gravitation works well in weak fields. Therefore, any creep effect around the Earth at a very short distance must be negligible. Therefore, the effects of dark matter must be insignificant on space-time around the Earth. This is what we will try to verify in this paragraph.

The value of the deformation of time in the case of the earth is first determined $\frac{c\delta t}{c\delta t_0}$

At r=20000 km from the earth, the GPS shift measured per day from 38 to 45 microseconds **[26].**

But the part due to strict gravitation effect is +45.9 Microseconds and -7 Microseconds for special relativity. So, we keep the gravity part to isolate the gravity effect alone:

$$\frac{\delta t}{\delta t_0} = \frac{45.9 \times 10^{-6}}{24 \times 60 \times 60} = 5.3125 \times 10^{-10}$$

We time dilation ($\varepsilon_{00,satellite}$) of the satellite level is:

$$\frac{GM}{rc^2} = \frac{6.674 \times 10^{-11} \times 5.972 \times 10^{24}}{26371000 \times 299792458^2} = 1.681166 \times 10^{-10}$$

At the surface of the Earth the time dilatation $\varepsilon_{00\,(Earth)}$ is:

$$\frac{GM}{rc^2} = \frac{6.674 \times 10^{-11} \times 5.972 \times 10^{24}}{6371000 \times 299792458^2} = 6.96077 \times 10^{-10}$$

So, the strain variation between the satellite level and the Earth surface is:

$$6.96077 \times 10^{-10} - 1.681166 \times 10^{-10} = 5.279604 \times 10^{-10}$$

Let us have a creep coefficient of: $\varphi_{time}$ :

$$\varphi_{time} = \left(\frac{5.3125 \times 10^{-10}}{5.27 \times 10^{-10}}\right) - 1 = 0.00632568$$

So un fact we obtain that the creep effect is quasi null around the Earth, that is normal because Newton Law works pretty well!

## 10. Discussion of the results

Thus, such creeps tend to show that space behaves like a granular medium. This overlaps the publications **[16]** and **[17]** on the transverse isotropy of space-time with regard to its behavior under gravitational waves.

Of course, this modification of the gravitational constant in $\frac{a_0}{a}$ is linked to the only MOND's theory for the constancy of the speed of stars in the galaxy. As it potentially also affects Einstein's constant $\kappa$, and as it is supposed to overcome the problem of the absence of detection of dark matter as a particle but that gravitational anomalies are very real, this modification should also be tested wherever the presence of dark matter is required (gravitational lensing, distribution of dark matter at the time of the big bang, etc) to be sure that this interpretation in terms of creep given by the elastic analogy is the right one. Additional tests should also be carried out for the temporal approach in large gravitational field to confirm this value for space of the creep coefficient φ of the order of magnitude between 0.2 and 9.

It is well known that dark matter does not have a small effect because in gravitation this unknown mass effect represents about 30% of the overall part that constitutes the universe. So, while we are trying to replace dark matter with a creep effect, this creep effect is also not minimal but corrects an important part of general relativity at the cosmological level.

From a more general point of view, this creep approach significantly modifies the Einstein's field equation which becomes:

$$R_{\mu\nu} + \frac{1}{2}g_{\mu\nu}R = \frac{8\pi G}{c^4}T_{\mu\nu} + \frac{8\pi G}{c^4}(1+\varphi)t_{e,\mu\nu} \quad (70)$$

$t_{e,\mu\nu}$ is the stress energy tensor connected with the space fabric itself as an elastic medium proposed by A. Tartaglia in **[37]** and M. Beau in **[38]**.

This additional stress-energy tensor $t_{\mu\nu}$ is also activated to explain the energy of the gravitation wave in vacuum **[36]**. The same approach is followed when we want to establish the energy of gravitational waves in a vacuum. The Einstein field equation becomes:

$$\boldsymbol{G}^{(1)}_{\mu\nu} = -\frac{8\pi G}{c^4}(\boldsymbol{T}_{\mu\nu} + \boldsymbol{t}_{\mu\nu}) \quad (71)$$

$G^{(1)}_{\mu\nu}$ is constructed from $G_{\mu\nu}$ terms of which are linear in $h_{\mu\nu}$ (see formula 4 of **[36]**):

$$t_{\mu\nu} = T^{GW}_{\mu\nu} = \frac{c^4}{8\pi G}\left[G^{(2)}_{\mu\nu} + \cdots\right] \quad (72)$$

$G^{(2)}_{\mu\nu}$ is constructed from quadratic terms of $h_{\mu\nu}$ (see formula 5 of **[36]**):

Or more in accordance with our approach of additional curvature (or amplification of the curvature due to the only visible mass/energy that does not change. The effects of dark matter (gravitational amplification) on vacuum are thus taken into account without the need for additional dark mass $\varphi > 0$). So, in vacuum we have:

$$\frac{1}{1+\varphi}\left\{R_{\mu\nu} + \frac{1}{2}g_{\mu\nu}R\right\} = \frac{8\pi G}{c^4} t_{e,\mu\nu} \quad (73a)$$

We verify well that as φ > 0 then 1/(1+φ) <1, so the curvature is reduced, so, the equivalent R radius increases and deflection too.

$$\left\{R_{\mu\nu} + \frac{1}{2}g_{\mu\nu}R\right\} = \frac{8\pi G}{c^4}(1+\varphi) t_{e,\mu\nu} \quad (73b)$$

As φ>0, we verify well that flexibility increases, so the rigidity decreases, and so the strains increase and the gravitation increases. $(1+\varphi)t_{e,\mu\nu}$ can be seen also as an additional dark matter that increases the gravitational effects.

## 11. Conclusion

We have therefore shown that within the framework of the elastic medium analogy, the effects of dark matter can be interpreted in the case of the anomaly of the star velocities around galaxies via the MOND theory or gravitational Lens as bullet cluster or from time dilatation data as a creep of the space medium.

This kind of behavior is typical of granular crystal **[17]** and therefore overlaps with other clues about the behavior of space:

- The regular repetition of the peaks in the case of X-ray diffractograms of lamellar crystal or clay in sheets as a function of the angles of refraction **[22]** and the peaks of the cosmic microwave background of the power spectra in temperature and polarizations when the photons passed through the Cosmic plasma at the time of the big bang giving the cosmic microwave background **[23]**),
- The great plasticity and creep effect of the crystal or granular medium under shear **[24]**, and their ability to self-clog comparable to the absence of residual tears in space-time behind rotating black holes
- The anisotropic Poisson coefficients 1 in one direction **[5]** and verry weak in the other.

Quantum gravity, even if it has not yet been validated by experiment, also leads to a quantification of space-time in small grains of the size of Planck. Our result is therefore consistent with the trend in current physics of space-time granularization. It remains to test this creep coefficient on the other physical phenomena where dark matter is involved to confirm or deny its value on the one hand and to verify whether we really need this dark matter to explain the discrepancy between the observed gravitation and the only visible masses that cause it or if, on the contrary, it is a decoy in the sense that space-time would flow under the only clearly visible masses. Finally, in the latter case, still according to the analogy of the elastic medium, space would behave like a crystal-type material generating deformations by creep and therefore gravitation in time under constant loading, i.e. a well-visible, classical and identified mass. But like any resolution of one mystery, another appears, if space-time is subject to creep, what is it constituted to have such a behavior? grains of space-time as proposed by loop quantum gravity? The question remains open.

Moreover, if dark matter is only a creep effect, time necessarily intervenes in the evolution of deformations. It is therefore necessary to study how dark matter has evolved over time. If it is stable, then it cannot be a creep effect. If, on the other hand, it varies and the gravitational effects increase like the distortions of space-time over time (a little like the small fluctuations in the density of the original plasma, we lead 13.7 billion years later to a cosmic web with bubbles of very large dimensions where on the surface there are galaxies and voids within the breasts of its bubbles) then the hypothesis of the creep of space-time instead of and the place of a dark matter that remains invisible to detection to this day remains credible. Time and experience will give us the answer if the creep coefficient for space $\varphi$ is effectively in the range +0.2/+9.

## Acknowledgement

We thank the reviewer for their judicious remarks which have made it possible to improve this article and especially the suggestion to use the reference **[21]** and **[44]** as a way to check the values of the creep coefficient proposed in this paper.